 \documentclass[preprint, superscriptaddress, preprintnumbers, amsmath, amssymb]{revtex4}

\allowdisplaybreaks
\allowdisplaybreaks[4]

\usepackage{CJK}
\usepackage{graphicx}
\usepackage{dcolumn}
\usepackage{bm}
\usepackage{multirow}
\usepackage[dvipdfm,
            pdfstartview=FitH,
            CJKbookmarks=true,
            bookmarksnumbered=true,
            bookmarksopen=true,
            colorlinks=true,
            pdfborder=001,
            linkcolor=blue,
            anchorcolor=blue,
            citecolor=blue
            ]{hyperref}

\begin{document}

\begin{CJK}{GBK}{song}

\title{Effective field theory for triaxially deformed odd-mass nuclei}

\author{Q. B. Chen}\email{qbchen@pku.edu.cn}
\affiliation{Physik-Department,
             Technische Universit\"{a}t M\"{u}nchen,
             D-85747 Garching, Germany}

\author{N. Kaiser}\email{nkaiser@ph.tum.de}
\affiliation{Physik-Department,
             Technische Universit\"{a}t M\"{u}nchen,
             D-85747 Garching, Germany}

\author{Ulf-G. Mei{\ss}ner}\email{meissner@hiskp.uni-bonn.de}
\affiliation{Helmholtz-Institut f\"{u}r Strahlen- und Kernphysik and
             Bethe Center for Theoretical Physics, Universit\"{a}t Bonn,
             D-53115 Bonn, Germany}
\affiliation{Institute for Advanced Simulation,
             Institut f\"{u}r Kernphysik and J\"{u}lich Center for Hadron Physics,
             Forschungszentrum J\"{u}lich,
             D-52425 J\"{u}lich, Germany}
\affiliation{Ivane Javakhishvili Tbilisi State University,
0186 Tbilisi, Georgia}

\author{J. Meng}\email{mengj@pku.edu.cn}
\affiliation{State Key Laboratory of Nuclear Physics and Technology,
             School of Physics, Peking University,
             Beijing 100871, China}
\affiliation{Yukawa Institute for Theoretical Physics,
             Kyoto University,
             Kyoto 606-8502, Japan}

\date{\today}

\begin{abstract}
  The effective field theory for collective rotations of triaxially deformed nuclei
  is generalized to odd-mass nuclei by including the angular momentum of the valence
  nucleon as an additional degree of freedom. The Hamiltonian is constructed up to
  next-to-leading order within the effective field theory formalism.
  The applicability of this Hamiltonian is examined by describing the wobbling
  bands observed in the lutetium isotopes $^{161,163,165,167}$Lu. It is found
  that by taking into account the next-to-leading order corrections, quartic
  in the rotor angular momentum, the wobbling energies $E_{\textrm{wob}}$ and
  spin-rotational frequency relations $\omega(I)$ are better described than
  with the leading order Hamiltonian.
\end{abstract}

\maketitle


\section{Introduction}

In Refs.~\cite{Q.B.Chen2017EPJA, Q.B.Chen2018PRC_v2}, an effective field
theory (EFT) has been established to describe the rotational and vibrational
motions of triaxially deformed even-even nuclei. The applicability of the
EFT Hamiltonian at leading order (LO) and next-to-leading order (NLO) has
been verified through a satisfactory description of the energy spectra
of ground state bands, $\gamma$-bands, and $K=4$ bands in the ruthenium
isotopes $^{102-112}$Ru.

EFT is an approach based on symmetry principles alone, and it exploits the
separation of scales for the systematic construction of the Hamiltonian
supplemented by a power counting. In this way, an increase in the
number of parameters (i.e., low-energy constants that need to be
adjusted to data) goes hand in hand with an increase in precision
and thereby counter balances the partial loss of predictive power.
Moreover, EFT often exhibits an impressive efficiency as highlighted
by analytical results and economical means of calculations. In
recent decades, chiral effective field theory has enjoyed
considerable success in low-energy hadronic and nuclear structure physics.
Pertinent examples include the descriptions of the nuclear
forces~\cite{Kolck1994PRC, Epelbaum2009RMP, Holt2019arXiv}, halo
nuclei~\cite{Bertulani2002NPA, Hammer2011NPA, Ryberg2014PRC},
and few-body systems~\cite{Bedaque2002ARNPS, Griesshammer2012PPNP,
Hammer2013RMP, Hammer2019arXiv}.

Papenbrock and collaborators have presented a series of works on an EFT for axially
deformed nuclei with the aim to investigate their rotational and vibrational
excitations~\cite{Papenbrock2011NPA, J.L.Zhang2013PRC, Papenbrock2014PRC,
Papenbrock2015JPG, Papenbrock2016PS, Perez2015PRC, Perez2015PRC_v1, Perez2016PRC}.
As mentioned at the beginning, we have extended this framework to triaxially
deformed even-even nuclei~\cite{Q.B.Chen2017EPJA, Q.B.Chen2018PRC_v2}.
In the present work, we will further generalize the EFT to odd-mass nuclei
with the aim to investigate their characteristic wobbling motion.

The wobbling motion, first proposed by Bohr and Mottelson in the
1970s~\cite{Bohr1975}, can occur only in a triaxially deformed nucleus
and hence becomes a unique fingerprint of the triaxial nuclear shape.
It manifests itself as an irregular precession of the rotational axis
around the axis with the largest moment of inertia (MoI). The energy
spectra related to this collective mode are called wobbling bands,
which consist of sequences of $\Delta I = 2$ rotational bands built
on different wobbling-phonon excitations~\cite{Bohr1975}.

For an odd-mass nucleus, the presence of the extra valence nucleon will
affect the wobbling mode. Depending on the relative orientation between the
quasi-particle angular momentum $\bm{j}$ and the (intermediate) axis
of the rotor with the largest MoI, two kinds of wobbling motion have been
proposed by Frauendorf and D\"{o}nau~\cite{Frauendorf2014PRC}.
If $\bm{j}$ is aligned parallel, one speaks of \emph{longitudinal}
wobbling, where the wobbling energy increases with the spin $I$. If the
alignment is perpendicular, this mode is called \emph{transverse}
wobbling, and the corresponding wobbling energy decreases with the
spin $I$.

Transverse wobbling bands have been reported in the mass region $A\approx 160$
for the isotopes $^{161}$Lu~\cite{Bringel2005EPJA}, $^{163}$Lu~\cite{Odegaard2001PRL,
Jensen2002PRL}, $^{165}$Lu~\cite{Schonwasser2003PLB}, $^{167}$Lu~\cite{Amro2003PLB},
and $^{167}$Ta~\cite{Hartley2009PRC}, in the mass region $A\approx 130$
for $^{135}$Pr~\cite{Matta2015PRL, Sensharma2019PLB}, and in the
mass region $A\approx 100$ for $^{105}$Pd~\cite{Timar2019PRL}. In this list,
$^{105}$Pd represents the first odd-neutron nucleus for which (transverse)
wobbling has been observed. In addition, transverse wobbling bands based on
a two-quasiparticle configuration have been reported and studied for the
even-even nucleus $^{130}$Ba~\cite{Petrache2019PLB, Q.B.Chen2019PRC_v1}.
On the other hand, longitudinal wobbling bands are rare and have been
observed up to now only in the isotopes $^{133}$La~\cite{Biswas2019EPJA}
and $^{187}$Au~\cite{Sensharma2020PRL}.

In this paper, we will first extend the EFT description of the collective
rotational motion of odd-mass nuclei by including the angular momentum of
the valence nucleon as a relevant degree of freedom. The obtained Hamiltonian
at NLO is applied to describe the wobbling bands in the lutetium isotopes
$^{161,163,165,167}$Lu. We also consider the corresponding inter-band
and intra-band electromagnetic transitions.


\section{Theoretical framework}\label{sec2}

In this section, the procedure of constructing the effective
Lagrangian and Hamiltonian for collective rotations of  triaxially
deformed odd-mass nuclei is outlined. It follows similar steps as
in the case of collective rotations of even-even triaxially deformed
nuclei investigated in Refs.~\cite{Q.B.Chen2017EPJA, Q.B.Chen2018PRC_v2}.

\subsection{Dynamical variables}

In an EFT, the symmetry is typically realized nonlinearly, and the
Nambu-Goldstone fields parametrize the coset space
$\mathcal{G}/\mathcal{H}$, where $\mathcal{G}$ is the symmetry group
of the Hamiltonian, and $\mathcal{H}$, the symmetry group of the
ground state, is a proper subgroup of $\mathcal{G}$~\cite{Coleman1969PR, Callan1969PR}.
The effective Lagrangian is built from those invariants that can be constructed
from the fields in the coset space. A triaxial
nucleus is invariant under $\textrm{D}_2=\textrm{Z}_2
\times \textrm{Z}_2$ (generated by rotations about the
body-fixed axes with an angle $\pi$), while SO(3) symmetry
is broken by the deformation. Hence, the Nambu-Goldstone
modes lie on the three-dimensional coset space
SO(3)/$\textrm{D}_2$~\cite{Q.B.Chen2017EPJA, Q.B.Chen2018PRC_v2}.

We write the Nambu-Goldstone fields in the body-fixed frame, where
the three generators of infinitesimal rotations about the body-fixed
$1$, $2$, and $3$-axes are $J_1$, $J_2$, and $J_3$. The modes depend
on three time-dependent Euler angles $\alpha(t)$, $\beta(t)$,
and $\gamma(t)$ that parametrize the unitary transformations
$U(\alpha,\beta,\gamma)$ related to SO(3) rotations in the following
way:
\begin{align}
 U(\alpha,\beta,\gamma)
 &=\exp\{-i\alpha(t)J_3\}\exp\{-i\beta(t)J_2\}
  \exp\{-i\gamma(t)J_3\}.
\end{align}
Note that the purely time-dependent variables $\alpha(t)$,
$\beta(t)$, and $\gamma(t)$ correspond to the zero modes of the
system. They  parametrize rotations of the deformed nucleus and
upon quantization they generate the rotational bands. Apparently,
one is dealing here with a field theory in zero space-dimensions,
i.e., ordinary quantum mechanics.

The underlying power counting is specified by
\begin{align}
 \alpha,\beta,\gamma \sim \mathcal{O}(1), \quad
 \dot{\alpha}, \dot{\beta}, \dot{\gamma} \sim \xi,
\end{align}
where the small parameter  $\xi$ denotes the energy scale of the
rotational motion and the dot refers to a time derivative.

\subsection{Building blocks}

The effective Lagrangian is built from invariants. These are
constructed from the components $a_t^1$, $a_t^2$, and $a_t^3$ of the
angular velocity arising from the decomposition
\begin{align}
 U^{-1}i\partial_t U=a_t^1J_1+a_t^2J_2+a_t^3J_3.
\end{align}
By taking appropriate traces of the matrix-exponentials, the
expansion coefficients read
\begin{align}
 a_t^1 &=-\dot{\alpha}\sin\beta\cos\gamma+\dot{\beta}\sin\gamma,\\
 a_t^2 &=\dot{\alpha}\sin\beta\sin\gamma+\dot{\beta}\cos\gamma,\\
 a_t^3 &=\dot{\alpha}\cos\beta+\dot{\gamma}.
\end{align}
One recognizes that these are the components of the angular velocity
of the nucleus in the body-fixed frame, according to rigid-body
kinematics~\cite{Landau1960book}.

\subsection{Effective Lagrangian}

Using the above building blocks, we have constructed in Ref.~\cite{Q.B.Chen2017EPJA}
the effective Lagrangian for the collective rotational motion of triaxially deformed
even-even nuclei at LO (up to $\xi^2$)
\begin{align}
  \mathcal{L}_{\textrm{LO}}^{\textrm{ee}}&=\frac{\mathcal{J}_1}{2}(a_t^1)^2
  +\frac{\mathcal{J}_2}{2}(a_t^2)^2+\frac{\mathcal{J}_3}{2}(a_t^3)^2,
\end{align}
and at NLO (up to $\xi^4$)
\begin{align}
  & \mathcal{L}_{\textrm{NLO}}^{\textrm{ee}}
  =\mathcal{L}_{\textrm{LO}}^{\textrm{ee}}
  +\frac{\mathcal{M}_1}{4}(a_t^1)^4
  +\frac{\mathcal{M}_2}{4}(a_t^2)^4+\frac{\mathcal{M}_3}{4}(a_t^3)^4\notag\\
  &\quad +\frac{\mathcal{M}_4}{2}(a_t^2)^2(a_t^3)^2+\frac{\mathcal{M}_5}{2}(a_t^1)^2(a_t^3)^2
  +\frac{\mathcal{M}_6}{2}(a_t^1)^2(a_t^2)^2,
\end{align}
with $\mathcal{J}_k$ ($k=1\ldots3$) and $\mathcal{M}_k$ ($k=1\ldots6$) the
parameters of moments of inertia to be determined from experimental data.

For an odd-mass nucleus, the extra degrees of freedom provided by the valence
nucleon must be included in the Lagrangian. Similar to Ref.~\cite{Papenbrock2011NPA},
we couple the angular momentum component $j_k$ of valence nucleon to the
Nambu-Goldstone modes $a_t^k$ as
\begin{align}
\mathcal{L}_{\textrm{C}}=j_1a_t^1+j_2a_t^2+j_3a_t^3,
\end{align}
which has the form of a Coriolis interaction~\cite{Bohr1975, Ring1980book}.

Hence, the corresponding LO and NLO Lagrangians for collective rotation of
odd-mass nuclei can be written as
\begin{align}
 \mathcal{L}_{\textrm{LO}}^{\textrm{eo}}
 &=\mathcal{L}_{\textrm{LO}}^{\textrm{ee}}+\mathcal{L}_{\textrm{C}},\\
 \mathcal{L}_{\textrm{NLO}}^{\textrm{eo}}
 &=\mathcal{L}_{\textrm{NLO}}^{\textrm{ee}}+\mathcal{L}_{\textrm{C}}.
\end{align}
Next, the corresponding Hamiltonians will be derived from these Lagrangians.

\subsection{Effective Hamiltonian}

We first derive the effective Hamiltonian at LO. From the LO Lagrangian
$\mathcal{L}_{\textrm{LO}}^{\textrm{eo}}$, one obtains the canonical momenta
as:
\begin{align}
\label{eq15}
 p_\alpha
 &=\frac{\partial \mathcal{L}_{\textrm{LO}}^{\textrm{eo}}}{\partial \dot{\alpha}}
 =-(a_t^1 \mathcal{J}_1+j_1)\sin\beta\cos\gamma
   +(a_t^2 \mathcal{J}_2+j_2)\sin\beta\sin\gamma \notag\\
 &\quad +(a_t^3 \mathcal{J}_3+j_3)\cos\beta,\\
\label{eq16}
 p_\beta
 &=\frac{\partial \mathcal{L}_{\textrm{LO}}^{\textrm{eo}}}{\partial \dot{\beta}}
  =(a_t^1 \mathcal{J}_1+j_1)\sin\gamma+(a_t^2 \mathcal{J}_2+j_2)\cos\gamma,\\
\label{eq17}
 p_\gamma
 &=\frac{\partial \mathcal{L}_{\textrm{LO}}^{\textrm{eo}}}{\partial \dot{\gamma}}
 =a_t^3 \mathcal{J}_3+j_3.
\end{align}
Obviously, by dropping $j_k$ in $a_t^k \mathcal{J}_k+j_k$, one gets back
the canonical momenta in the case of an even-even nucleus written in Eqs.~(8)-(10)
of Ref.~\cite{Q.B.Chen2017EPJA}.

Using a Legendre transformation, the Hamiltonian is obtained as
\begin{align}
  \mathcal{H}_{\textrm{LO}}^{\textrm{eo}}
  &=\dot{\alpha}p_\alpha+\dot{\beta}p_\beta+\dot{\gamma}p_\gamma
  -\mathcal{L}_{\textrm{LO}}\notag\\
  &=\frac{1}{2\mathcal{J}_1}\Big[\Big(-\frac{\cos\gamma}{\sin\beta}p_\alpha+\sin\gamma p_\beta
  +\cos\gamma\cot\beta p_\gamma\Big)-j_1 \Big]^2\notag\\
 &+\frac{1}{2\mathcal{J}_2}\Big[\Big(\frac{\sin\gamma}{\sin\beta}p_\alpha+\cos\gamma p_\beta
  -\sin\gamma\cot\beta p_\gamma\Big)-j_2 \Big]^2\notag\\
 &+\frac{1}{2\mathcal{J}_3}\Big[(p_\gamma)-j_3\Big]^2.
\end{align}
Note that the expressions in the round brackets are the three components of the
total angular momentum $I_1$, $I_2$, and $I_3$ with respect to the body-fixed
frame~\cite{Ring1980book},
\begin{align}
 I_1&=-\frac{\cos\gamma}{\sin\beta} p_\alpha + \sin\gamma p_\beta+
  \cos\gamma\cot\beta p_\gamma,\\
 I_2&=\frac{\sin\gamma}{\sin\beta} p_\alpha + \cos\gamma p_\beta -
  \sin\gamma\cot\beta p_\gamma,\\
 I_3&=p_\gamma.
\end{align}
Consequently, the Hamiltonian at LO reads
\begin{align}
  \mathcal{H}_{\textrm{LO}}^{\textrm{eo}}=\frac{(I_1-j_1)^2}{2\mathcal{J}_1}
 +\frac{(I_2-j_2)^2}{2\mathcal{J}_2}+\frac{(I_3-j_3)^2}{2\mathcal{J}_3},
\end{align}
and this formula represents the Hamiltonian of a triaxial rotor with
angular momentum components $R_k=I_k-j_k$. Obviously, the rotor and
quasi-particle angular momenta are coupled to the total angular momentum
by $\bm{I}=\bm{R}+\bm{j}$.

Following a similar procedure, one derives the effective Hamiltonian at
NLO as
\begin{align}
  \mathcal{H}_{\textrm{NLO}}^{\textrm{eo}}
  &=\mathcal{H}_{\textrm{LO}}^{\textrm{eo}}\notag\\
  &-\frac{\mathcal{M}_1 (I_1-j_1)^4}{4\mathcal{J}_1^4}
   -\frac{\mathcal{M}_2 (I_2-j_2)^4}{4\mathcal{J}_2^4}
   -\frac{\mathcal{M}_3 (I_3-j_3)^4}{4\mathcal{J}_3^4}\notag\\
  &-\frac{\mathcal{M}_4 [(I_2-j_2)^2(I_3-j_3)^2+(I_3-j_3)^2(I_2-j_2)^2]}{4\mathcal{J}_2^2\mathcal{J}_3^2}\notag\\
  &-\frac{\mathcal{M}_5 [(I_3-j_3)^2(I_1-j_1)^2+(I_1-j_1)^2(I_3-j_3)^2]}{4\mathcal{J}_1^2\mathcal{J}_3^2}\notag\\
  &-\frac{\mathcal{M}_6 [(I_1-j_1)^2(I_2-j_2)^2+(I_2-j_2)^2(I_1-j_1)^2]}{4\mathcal{J}_1^2\mathcal{J}_2^2}.
\end{align}
Note that the last three terms have been written as anticommutators to guarantee
a hermitean Hamiltonian. Again, by dropping the $j_k$ one recovers the NLO effective
Hamiltonian of the even-even rotor given in Eq.~(36) of Ref.~\cite{Q.B.Chen2017EPJA}.

At this point, one needs also a Hamiltonian that describes the motion
of the valence nucleon. Since it cannot be derived from the EFT concept based on
Nambu-Goldstone modes, we borrow it from the phenomenologically successful Nilsson
model in the form of a single-$j$ shell Hamiltonian~\cite{Bohr1975, Ring1980book}
\begin{align}\label{eq18}
h_{\textrm{p}}=\frac{C}{2}\Big\{\cos \gamma_2\Big[j_3^2-\frac{j(j+1)}{3}\Big]
 +\frac{\sin\gamma_2}{2\sqrt{3}}(j_+^2+j_-^2)\Big\},
\end{align}
with $\gamma_2$ the triaxial deformation parameter. The coupling parameter
$C$ is related to the axial deformation parameter $\beta_2$ by~\cite{S.Y.Wang2009CPL}
\begin{align}
 C=\frac{123}{8}\sqrt{\frac{5}{\pi}}\frac{2N+3}{j(j+1)}A^{-1/3}\beta_2,
\end{align}
with $N$ the oscillator quantum number of the major shell embedding the
single-$j$ shell, and $A$ the mass number.

With this completion, the total rotational Hamiltonian for odd-mass
nuclei reads
\begin{align}
 \mathcal{H}_{\textrm{LO}} &=h_{\textrm{p}}+\mathcal{H}_{\textrm{LO}}^{\textrm{eo}},\\
 \mathcal{H}_{\textrm{NLO}}&=h_{\textrm{p}}+\mathcal{H}_{\textrm{NLO}}^{\textrm{eo}}.
\end{align}
Note that $\mathcal{H}_{\textrm{LO}}$ is nothing but the renowned
particle-rotor model Hamiltonian~\cite{Bohr1975, Ring1980book}.

\subsection{Solutions of the rotational Hamiltonian}

The Hamiltonian can be solved by diagonalization in a complete basis.
Here, we adopt the so-called weak-coupling basis~\cite{Bohr1975, Ring1980book,
Streck2018PRC, Q.B.Chen2019PRC}
\begin{align}\label{eq5}
 |IMjRK_R\rangle=\sum_{m,M_R}\langle j m R M_R|IM\rangle~|jm\rangle
 \otimes |RM_R K_R \rangle,
\end{align}
where $I$ denotes the total angular momentum quantum number of
the odd-mass nuclear system (rotor plus particle). Furthermore,
$m$, $M_R$, and $M$ are the quantum numbers corresponding to the
projections of $\bm{j}$, $\bm{R}$, and $\bm{I}$ onto the 3-axis of the
laboratory frame, and $K_R$ is related to the projection of $\bm{R}$
onto the 3-axis of the principal axes frame. Obviously, the appearance of
Clebsch-Gordan coefficients $\langle j m R M_R|IM\rangle$ requires
$M=m+M_R$, and the values of $R$ must satisfy the triangular condition
$|I-j| \leq R \leq I+j$ of angular momentum coupling.

Making use of Wigner-functions, the rotational wave functions of the
particle and the rotor in Eq.~(\ref{eq5}) can be written as
\begin{align}
\label{eq3}
 |jm\rangle &=\sum_{\Omega=-j}^j D_{m\Omega}^j|j\Omega\rangle,\\
\label{eq4}
 |RM_R K_R\rangle &=\sqrt{\frac{2R+1}{16\pi^2(1+\delta_{K_R0})}}
  \Big[D_{M_RK_R}^R +(-1)^R D_{M_R-K_R}^R \Big],
\end{align}
where $\Omega$ is the quantum number related to the 3-axis component
of the particle angular momentum $\bm{j}$ in the intrinsic frame.
Furthermore, $K_R$ is an even integer ranging from $0$ to $R$,
where $K_R=0$ is excluded for odd $R$. Both restrictions on $K_R$
come from the $\textrm{D}_2$ symmetry of a triaxial nucleus~\cite{Bohr1975}.
Note that for an axially symmetric nucleus, $R$ can take only
even integer values since $K_R$ must be zero.

The matrix elements of the collective rotor Hamiltonian can now be
calculated easily as
\begin{align}
 \langle IMjRK_R^\prime|\mathcal{H}_{\textrm{LO/NLO}}^{\textrm{eo}}|IMjRK_R \rangle=
 \sum_i c_{K_R^\prime}^{Ri} E_{Ri} c_{K_R}^{Ri},
\end{align}
where the eigenenergies $E_{Ri}$ and corresponding coefficients
$c_{K_R}^{Ri}$ ($i$ labels the different eigenstates) of the eigenvectors
are obtained by diagonalizing the collective rotor Hamiltonian
$\mathcal{H}_{\textrm{LO/NLO}}^{\textrm{eo}}$ in the basis $|RM_RK_R\rangle$ introduced
in Eq.~(\ref{eq4}):
\begin{align}
 \mathcal{H}_{\textrm{LO/NLO}}^{\textrm{eo}}|RM_Ri\rangle &= E_{Ri}|RM_Ri\rangle,\\
 |RM_Ri\rangle &=\sum_{K_R} c_{K_R}^{Ri} |RM_RK_R\rangle.
\end{align}

\subsection{Electromagnetic transitions}

In general, the electromagnetic transition probability is calculated
as~\cite{Bohr1975, Ring1980book}
\begin{align}
 &\quad B(\sigma\lambda,I^\prime \to I)=\frac{1}{2I^\prime +1}\sum_{\mu M^\prime}
  \big|\langle IM|\mathcal{M}(\sigma\lambda, \mu)|I^\prime M^\prime\rangle \big|^2,
\end{align}
where $\mathcal{M}(\sigma\lambda, \mu)$ is the electromagnetic
transition operator of multi-polarity $\sigma\lambda$, and
$M=M^\prime+\mu$.

In the present work, the rotational wave function $|IM\rangle$ is expanded
in the weak-coupling basis (\ref{eq5}) as
\begin{align}
  & |IM\rangle =\sum_{R,K_R} c_{R,K_R}|IMjRK_R\rangle
  =\sum_{R,K_R,m} d_{m, R,K_R} |jm\rangle|RM_RK_R\rangle,
\end{align}
with $d_{m,R,K_R}=c_{R,K_R} \langle j m R M_R|IM\rangle$. Therefore,
the electromagnetic transition probability can be rewritten as
\begin{align}\label{eq19}
  B(\sigma\lambda,I^\prime \to I)
  & =\frac{1}{2I^\prime +1}\sum_{\mu M^\prime}\Big|
    \sum_{R^\prime,K_R^\prime, m^\prime}
    \sum_{R,K_R, m} d_{m, R,K_R}
    d_{m^\prime, R^\prime,K_R^\prime}\notag\\
   &\quad \times \langle j m | \langle R M_R K_R|
     \mathcal{M}(\sigma\lambda, \mu) |j m^\prime \rangle
     |R^\prime M_R^\prime K_R^\prime \rangle\Big|^2.
\end{align}

For an $E2$ transition, the corresponding operator is
\begin{align}
 \mathcal{M}(E2,\mu)=\sqrt{\frac{5}{16\pi}}Q_{2\mu},
\end{align}
with the quadrupole moments in the laboratory frame
\begin{align}
 Q_{2\mu}=\sum_{\nu}D_{\mu\nu}^{2*}Q_{2\nu}^\prime,
\end{align}
obtained by a rotation from the quadrupole moments in the
principal axis frame
\begin{align}
 &Q_{20}^\prime=Q_0\cos\gamma_2, \notag\\
 &Q_{21}^\prime=Q_{2-1}^\prime=0,\notag\\
 &Q_{22}^\prime=Q_{2-2}^\prime=\frac{1}{\sqrt{2}}Q_0\sin\gamma_2.
\end{align}
Here, $Q_0=(3/\sqrt{5\pi})R_0^2Z\beta_2$ is the intrinsic quadrupole
moment, with $R_0=1.2A^{1/3}$ the nuclear radius, and $Z$ the charge number.
Since $m^\prime=m$ is required in Eq.~(\ref{eq19}) for $E2$ transition,
one gets
\begin{align}
  B(E2,I^\prime \to I)
   &=\frac{1}{2I^\prime +1}\sum_{\mu M^\prime}\Big|
    \sum_{R,K_R, R^\prime,K_R^\prime, m} d_{m, R,K_R}
    d_{m, R^\prime,K_R^\prime}\notag\\
   &\quad \times \langle R M_R K_R|
     \mathcal{M}(E2, \mu)
     |R^\prime M_R^\prime K_R^\prime \rangle\Big|^2.
\end{align}

For a $M1$ transition, the operator $\mathcal{M}(M1,\mu)$ is composed
of the particle and rotor angular momenta in the laboratory frame
\begin{align}\label{eq20}
 \mathcal{M}(M1,\mu)
  &=g_p\hat{j}_{\mu}+ g_R \hat{R}_\mu\notag\\
  &=g_R\hat{I}_\mu+(g_p-g_R)\hat{j}_{\mu},
\end{align}
where $g_p$ and $g_R$ are the $g$-factors of particle and rotor. Note that
since transition matrix elements of $\hat{I}_\mu$ vanish, the first term
can be dropped. In the second term, $\hat{j}_\mu$ with $\mu=0,\pm 1$ denotes
the spherical components of the particle angular momentum,
\begin{align}
 \hat{j}_0 =j_3, \quad
 \hat{j}_{\pm 1} =\mp \frac{1}{\sqrt{2}}(j_1 \pm ij_2).
\end{align}
Now, one has in Eq.~(\ref{eq19}) the conditions $R^\prime=R$, $M_R^\prime=M_R$,
and $K_R^\prime=K_R$, and therefore the $M1$ transition probability
simplifies to
\begin{align}
  B(M1,I^\prime \to I)
   &=\frac{1}{2I^\prime +1}\sum_{\mu M^\prime}\Big|
    \sum_{R,K_R, m, m^\prime} d_{m, R,K_R}
    d_{m^\prime, R^\prime,K_R^\prime}\notag\\
   &\quad \times \langle j m | \mathcal{M}(M1, \mu)
   |j m^\prime \rangle \Big|^2.
\end{align}


\section{Results and discussion}

In the calculations, the moment of inertia parameters $\mathcal{J}_k$ and
$\mathcal{M}_k$ appearing in the LO and NLO Hamiltonian are fitted to the data.
The deformation parameters for the single-$j$ shell Hamiltonian (\ref{eq18})
are taken from Ref.~\cite{Odegaard2001PRL} as $\beta_2=0.40$ and $\gamma_2=20^\circ$.

\subsection{Wobbling bands in $^{163}$Lu}

\begin{figure}[!ht]
  \begin{center}
    \includegraphics[width=8.5 cm]{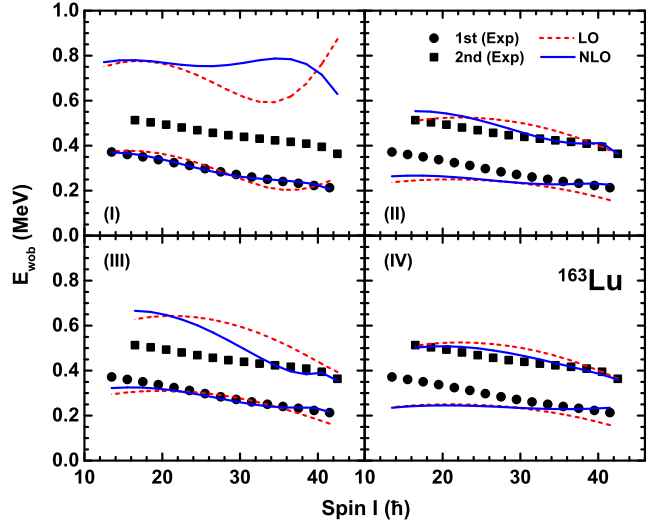}
    \caption{(Color online) Wobbling energies of the first and second wobbling
    bands as functions of spin $I$ in $^{163}$Lu calculated at LO and
    NLO by four fitting strategies (I)-(IV) in comparison to the experimental
    data~\cite{Odegaard2001PRL, Jensen2002PRL}.}\label{fig3}
  \end{center}
\end{figure}

\begin{figure}[!ht]
  \begin{center}
    \includegraphics[width=8.5 cm]{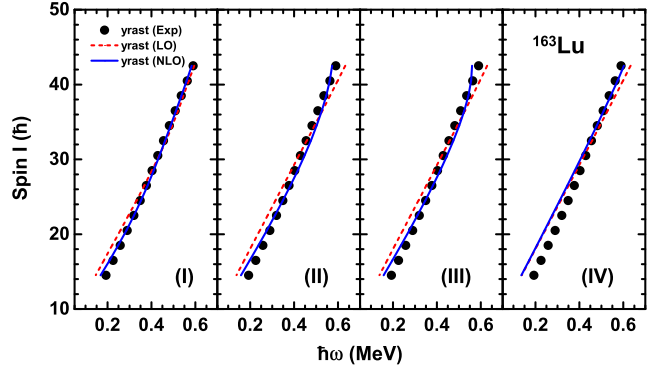}
    \caption{(Color online) Spin-rotational frequency relationships for
    the yrast band in $^{163}$Lu calculated at LO and NLO by four fitting
    strategies (I)-(IV) in comparison with experimental
    data~\cite{Odegaard2001PRL, Jensen2002PRL}.}\label{fig4}
  \end{center}
\end{figure}

In Fig.~\ref{fig3}, the energy differences with respect to the yrast band
(called wobbling energies $E_{\textrm{wob}}$) of the first and second wobbling bands
in $^{163}$Lu as calculated in the EFT at LO and NLO are shown in comparison to the
experimental data~\cite{Odegaard2001PRL, Jensen2002PRL}. In Fig.~\ref{fig4}, the
spin-rotational frequency relationships $\omega(I) =[E(I)-E(I-2)]/2$ for the yrast
band in $^{163}$Lu as calculated in the EFT at LO and NLO are shown in comparison
to the experimental data~\cite{Odegaard2001PRL, Jensen2002PRL}. For the various fits,
the obtained parameters are listed in Table~\ref{tab1}.

In our EFT calculations, we have adopted four strategies of fitting (I)-(IV).
In fit (I), the wobbling energies of the first wobbling band and
the rotational frequencies of the yrast band are used to determine simultaneously
the parameters $\mathcal{J}_k$ and $\mathcal{M}_k$ appearing in the EFT
formalism. It is seen that the wobbling energies of the first wobbling
band as well as the rotational frequencies of the yrast band
can be reproduced. But the calculated wobbling energies of the
second wobbling band are overpredicted by about 0.3 MeV with respect
to the experimental values.

In fit (II), we have further included the wobbling energies of the
second wobbling band to determine the parameters $\mathcal{J}_k$ and
$\mathcal{M}_k$. In this case, the description of the second wobbling
band gets much improved. In particular, the calculated wobbling energies
of the second wobbling band decrease with spin $I$. However, the
wobbling energy of the first wobbling band is underestimated about
0.1 MeV in the low spin region.

\begin{table}[!ht]
\caption{Parameters used in the LO and NLO calculations in four fitting
strategies for $^{163}$Lu. The units of $\mathcal{J}_k$ and $\mathcal{M}_k$
are $\hbar^2/\textrm{MeV}$ and $\hbar^4/\textrm{MeV}^3$, respectively.} \label{tab1}
\begin{ruledtabular}
\begin{tabular}{cccccccccccc}
 & Nucleus    & (I) & (II) & (III) & (IV) \\
\colrule
\multirow{3}{*}{LO}
    & $\mathcal{J}_1$ &    65.48    &  61.67      &   62.88      &   61.67  \\
    & $\mathcal{J}_2$ &    50.51    &  55.46      &   55.02      &   55.46  \\
    & $\mathcal{J}_3$ &     4.09    &  12.56      &    7.89      &   12.56  \\
\hline
\multirow{9}{*}{NLO}
    & $\mathcal{J}_1$ &    60.56    &  53.18      &   54.24      &   61.67  \\
    & $\mathcal{J}_2$ &    44.45    &  47.13      &   47.08      &   56.46  \\
    & $\mathcal{J}_3$ &     3.85    &  10.43      &    6.87      &   12.56  \\
    & $\mathcal{M}_1$ &   $-4.93$   &  19.99      &   19.99      &   18.44  \\
    & $\mathcal{M}_2$ &    19.77    &  19.07      &   19.04      &    7.73  \\
    & $\mathcal{M}_3$ &   $-0.19$   &   0.55      &  $-0.41$     &    1.43  \\
    & $\mathcal{M}_4$ &  $-19.97$   &  12.46      &    10.12     &   15.70  \\
    & $\mathcal{M}_5$ &     9.94    &  16.45      &    14.29     &  $-3.19$  \\
    & $\mathcal{M}_6$ &    19.99    &   6.21      &     3.71     & $-14.25$  \\
\end{tabular}
\end{ruledtabular}
\end{table}

To address the problem appearing in fit (II), we have enlarged in the total
$\chi^2$ the weight of the first wobbling band by a factor 10. This fitting
strategy (III) can indeed improve the description of the first wobbling band as
shown in Fig.~\ref{fig3}. However, the wobbling energies of the second wobbling
band are overestimated again in the low spin region.

In fitting strategy (IV), we take the full data collection as in fitting
strategy (II), but keep fixed the LO parameters $\mathcal{J}_k$ when performing
the fit of $\mathcal{M}_k$ at NLO. As can be seen, the description of the
wobbling energies does not change much. However, as shown in Fig.~\ref{fig4},
the agreement for the spin-rotational frequency relationships of the yrast
band becomes a bit worse.

\subsection{Wobbling bands in $^{161,165,167}$Lu}

\begin{figure}[!ht]
  \begin{center}
    \includegraphics[width=5.5 cm]{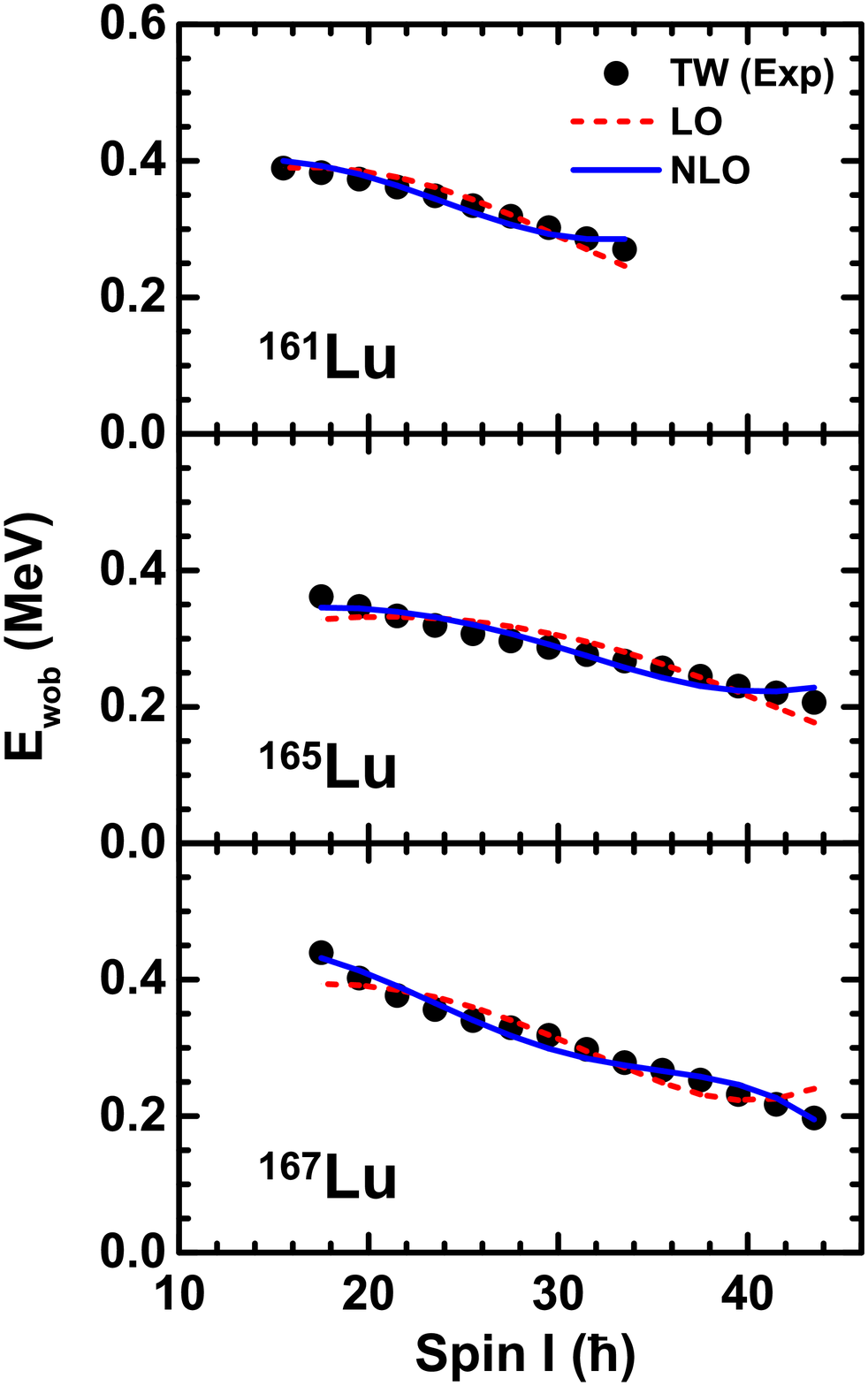}
    \caption{(Color online) Wobbling energy as functions of spin in
    $^{161,165,167}$Lu as calculated at LO and NLO in comparison
    with experimental data~\cite{Bringel2005EPJA, Schonwasser2003PLB,
    Amro2003PLB}.}\label{fig1}
  \end{center}
\end{figure}

\begin{figure}[!ht]
  \begin{center}
    \includegraphics[width=7.7 cm]{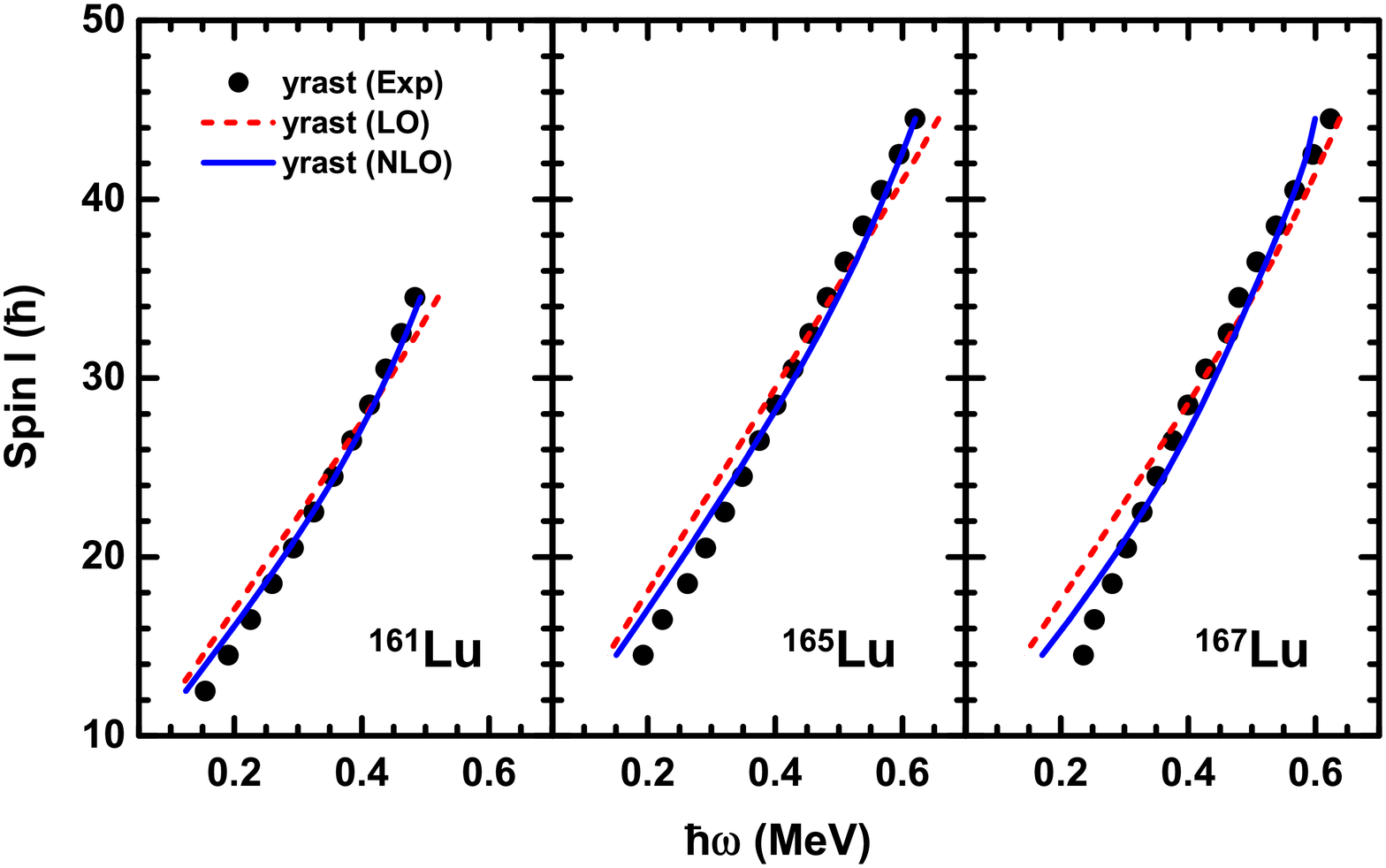}\\
    \includegraphics[width=7.7 cm]{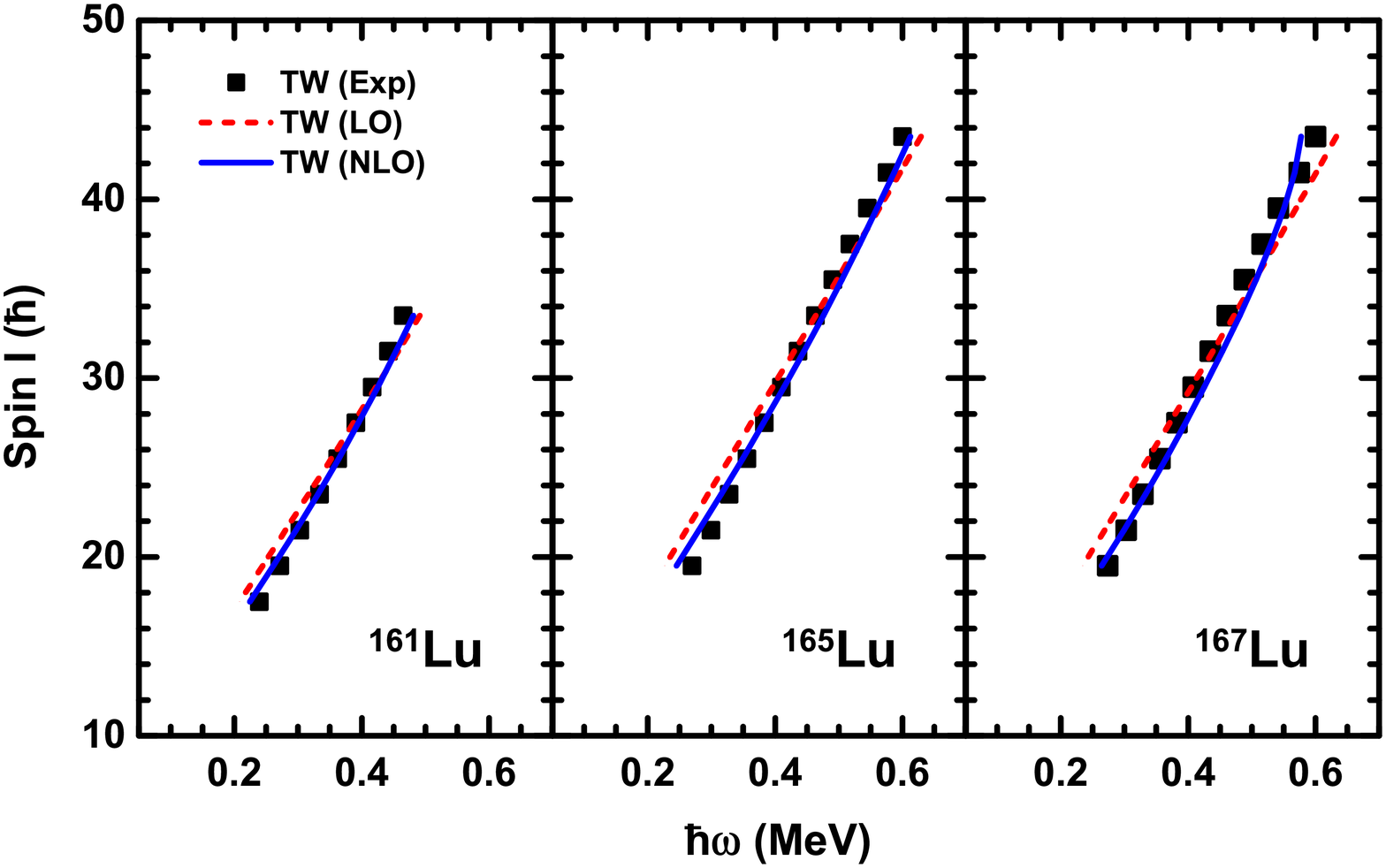}
    \caption{(Color online) Spin-rotational frequency relationships for
    the yrast and wobbling bands in $^{161,165,167}$Lu as calculated
    at LO and NLO in comparison with the experimental
    data~\cite{Bringel2005EPJA, Schonwasser2003PLB, Amro2003PLB}.}\label{fig2}
  \end{center}
\end{figure}

In Fig.~\ref{fig1}, the wobbling energies as functions of spin $I$
in $^{161,165,167}$Lu as calculated at LO and NLO are shown in comparison
with experimental data~\cite{Bringel2005EPJA, Schonwasser2003PLB, Amro2003PLB}.
The results of spin-rotational frequency relationships for their yrast and wobbling
bands are displayed in Fig.~\ref{fig2}. In the EFT calculations, the experimental
data of the wobbling energies of the first wobbling band and the rotational
frequencies of the yrast band are used to determine the parameters $\mathcal{J}_k$
and $\mathcal{M}_k$. The obtained best-fit parameters for each Lu isotope
are given in Table~\ref{tab2}.

From Figs.~\ref{fig1} and \ref{fig2}, one can see that the EFT at LO and NLO can
both reproduce this large amount of experimental data. In particular, the good
agreement for the spin-rotational frequency relationships of the second wobbling
band indicates that the EFT is quite efficient. The wobbling energies of these
three nuclei decrease with spin $I$, thus exhibiting the feature of transverse
wobbling. The descriptions at NLO are generally better than LO, which points to the
relevance of the higher-order terms in the NLO Hamiltonian. In particular,
the LO calculation shows an earlier termination of the transverse wobbling
motion in $^{167}$Lu. In addition, the LO calculation gives much smaller
rotational frequencies at low spins, i.e., it overestimates the dynamical
moment of inertia $dI/d\omega$.

\begin{table}[!ht]
\caption{Parameters used in the LO and NLO calculations for $^{161,165,167}$Lu.
The units of $\mathcal{J}_k$ and $\mathcal{M}_k$ are $\hbar^2/\textrm{MeV}$ and
$\hbar^4/\textrm{MeV}^3$, respectively.} \label{tab2}
\begin{ruledtabular}
\begin{tabular}{cccccccccccc}
 & Nucleus    & $^{161}$Lu & $^{165}$Lu & $^{167}$Lu \\
\colrule
\multirow{3}{*}{LO}
    & $\mathcal{J}_1$ &   61.89    &  63.11      &   64.10   \\
    & $\mathcal{J}_2$ &   49.53    &  55.52      &   52.10   \\
    & $\mathcal{J}_3$ &    4.23    &   7.05      &    4.20   \\
\hline
\multirow{9}{*}{NLO}
    & $\mathcal{J}_1$ &   60.59    &  60.90      &   59.66   \\
    & $\mathcal{J}_2$ &   44.61    &  50.04      &   43.17   \\
    & $\mathcal{J}_3$ &    3.38    &   5.66      &    2.64   \\
    & $\mathcal{M}_1$ &    9.99    &   9.96      & $-10.00$  \\
    & $\mathcal{M}_2$ &    9.99    &   9.99      &   10.00   \\
    & $\mathcal{M}_3$ &    0.05    &   0.14      &  $-0.38$  \\
    & $\mathcal{M}_4$ &  $-9.99$   & $-9.91$     &  $-9.99$  \\
    & $\mathcal{M}_5$ &    3.03    &   0.92      &    8.92   \\
    & $\mathcal{M}_6$ &    9.99    &   9.98      &    9.99   \\
\end{tabular}
\end{ruledtabular}
\end{table}

\subsection{Electromagnetic transitions}

\begin{figure}[!ht]
  \begin{center}
    \includegraphics[width=7.7 cm]{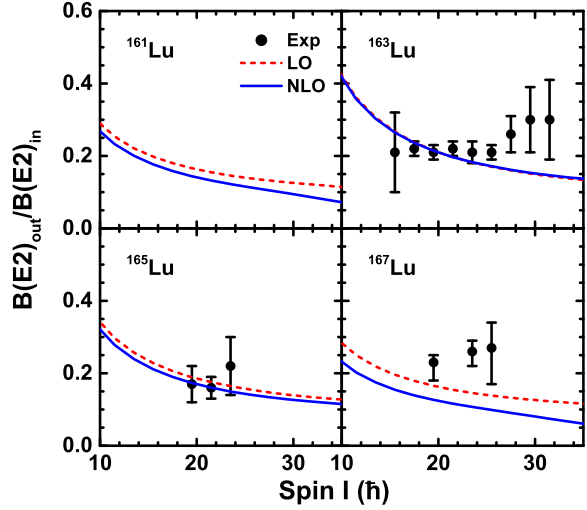}
    \caption{(Color online) The $B(E2)_{\textrm{out}}/B(E2)_{\textrm{in}}$ ratio
    for the transitions from wobbling to yrast band in $^{161,163, 165,167}$Lu
    as calculated at LO and NLO in comparison with the available experimental
    data~\cite{Odegaard2001PRL, Jensen2002PRL, Schonwasser2003PLB, Amro2003PLB}.}\label{fig5}
  \end{center}
\end{figure}

\begin{figure}[!ht]
  \begin{center}
    \includegraphics[width=7.7 cm]{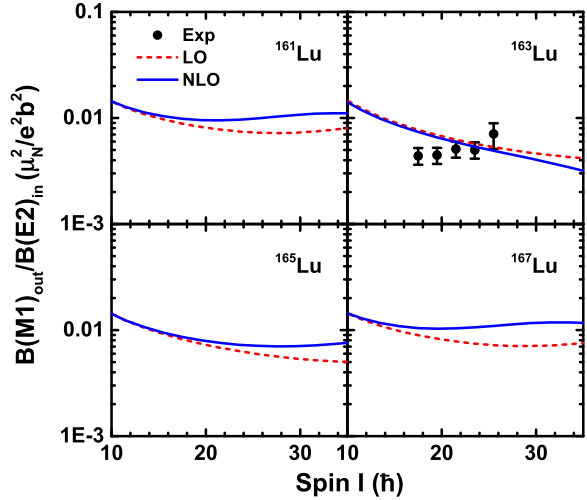}
    \caption{(Color online) The $B(M1)_{\textrm{out}}/B(E2)_{\textrm{in}}$ ratio
    for the transitions from wobbling to yrast band in $^{161,163, 165,167}$Lu
    as calculated at LO and NLO in comparison with the available
    experimental data~\cite{Odegaard2001PRL, Jensen2002PRL}.}\label{fig6}
  \end{center}
\end{figure}

A hallmark of the wobbling mode on electromagnetic transition
properties is the enhancement of the electric quadrupole component
for $\Delta I=1$ transitions between the neighboring
wobbling bands. In Figs.~\ref{fig5} and \ref{fig6}, the
$B(E2)_{\textrm{out}}/B(E2)_{\textrm{in}}$ and $B(M1)_{\textrm{out}}
/B(E2)_{\textrm{in}}$ ratios for the transitions from the first
wobbling band to the yrast band in $^{161,163, 165,167}$Lu as
calculated at LO and NLO are shown in comparison with the available
experimental data~\cite{Odegaard2001PRL, Schonwasser2003PLB, Amro2003PLB}.
Here, ``out'' denotes the $\Delta I=1$ inter-band transitions (wobbling
$\to$ yrast), and ``in'' refers to $\Delta I=2$ intra-band
transitions within the same wobbling band. Note that the results for
$^{163}$Lu have been obtained by fitting strategy (II), which
gives the best agreement with the experimental wobbling energies
and spin-rotational frequency relationships as shown in Figs.~\ref{fig3}
and \ref{fig4}.

The large $B(E2)_{\textrm{out}}/B(E2)_{\textrm{in}}$ values and
very small $B(M1)_{\textrm{out}}/B(E2)_{\textrm{in}}$ values support the
occurrence of transverse wobbling motion in the Lu isotopes. Note that most
of the experimental $B(E2)_{\textrm{out}}/B(E2)_{\textrm{in}}$ values can be
reproduced within error bars by the EFT both at LO and NLO.

In order to reproduce qualitatively the experimental values
of $B(M1)_{\textrm{out}}/B(E2)_{\textrm{in}}$, we have quenched
the relevant $g$-factor $(g_p-g_R)$ in Eq.~(\ref{eq20}) by a factor 0.25.
According to Ref.~\cite{Frauendorf2015PRC}, such a quenching factor allows to
take into account effects of the scissor mode, which mixes with the wobbling mode
and could reduce the $B(M1)_{\textrm{out}}$ by a factor of 3-20.

\section{Summary}

The effective field theory for the collective motion of triaxially deformed nuclei
has been generalized to odd-mass nuclei by including the angular momentum of the
valence nucleon as an additional degree of freedom. The Hamiltonian has been
constructed up to next-to-leading order, where it goes beyond the particle-rotor
model by quartic terms. We have examined its applicability by calculating the
wobbling bands in the lutetium isotopes $^{161,163,165,167}$Lu. It is found
that by taking into account the NLO order corrections, the experimental wobbling
energies and spin-rotational frequency relations are better described than at LO
Hamiltonian, which points to the relevance of the higher-order terms in
the NLO Hamiltonian. At the same time, the electromagnetic transition
strengths for inter-band and intra-band transitions are well described.

Next, we will further generalize the EFT by including the coupling
of a proton and a neutron to the rotor and apply it to investigate
chiral doublet bands~\cite{Frauendorf1997NPA}.

\section*{Acknowledgements}

This work has been supported in parts by Deutsche Forschungsgemeinschaft (DFG)
and National Natural Science Foundation of China (NSFC) through funds provided
by the Sino-German CRC 110 ``Symmetries and the Emergence of Structure in QCD''
(DFG Grant No. TRR110 and NSFC Grant No. 11621131001), the
National Key R\&D Program of China (Contract No. 2017YFE0116700 and
No. 2018YFA0404400), and the NSFC under Grant No. 11935003. The work of U.-G.M.
was also supported by the Chinese Academy of Sciences (CAS) through a President's
International Fellowship Initiative (PIFI) (Grant No. 2018DM0034) and by the
VolkswagenStiftung (Grant No. 93562).


\end{CJK}

\end{document}